\documentclass{iau}
\usepackage{graphicx}
\usepackage{graphics}
\usepackage{bm}

\title[Magnetic fields - shaping stars and planets] 
{Stellar magnetic activity and their influence on the habitability of exoplanets}

\author[T. L\"uftinger, M. G\"udel, \& C. Johnstone]   
{T. L\"uftinger$^1$, M. G\"udel$^1$ and C. Johnstone$^1$ 
}

\affiliation{
$^1$Dept. of Astronomy, University of Vienna, T\"urkenschanzstr. 17, A-1180 Vienna, Austria \\ e-mail: {\tt theresa.rank-lueftinger@univie.ac.at, manuel.guedel@univie.ac.at, colin.johnstone@univie.ac.at} 
} 

\pubyear{2015}
\volume{305}  
\jname{Polarimetry: From the Sun to Stars and Stellar Environments}
\editors{K.N. Nagendra, S. Bagnulo, \\ R. Centeno, \& M. J. Mart\' inez Gonz\'alez, eds.}
\begin{document}

\maketitle

\begin{abstract}
Stellar magnetism, explorable via polarimetry, is a crucial driver of activity, ionization, photodissociation,
chemistry and winds in stellar environments. Thus it has an important impact on the atmospheres and 
magnetospheres of surrounding planets. Modeling of stellar magnetic fields and their winds is extremely
challenging, both from the observational and the theoretical points of view, and only recent ground
breaking advances in observational instrumentation - as were discussed during this Symposium - and a 
deeper theoretical understanding of magnetohydrodynamic processes in stars enable us to model stellar
magnetic fields and winds and the resulting influence on surrounding planets in more and more detail.
We have initiated a national and international research network (NFN): 'Pathways to Habitability - From
Disks to Active Stars, Planets to Life', to address questions on the formation and habitability of
environments in young, active stellar/planetary systems. In this contribution we discuss the work we are
carrying out within this project and focus on how stellar magnetic fields, their winds and the relation
to stellar rotation can be assessed observationally with relevant techniques such as Zeeman Doppler
Imaging (ZDI), field extrapolation and wind simulations. 
\end{abstract}

\firstsection 

\section{Introduction}
\label{lueftinger-sec-intro}

Stellar magnetic fields play a key role in the the formation of stars, their structure and evolution and in the generation of winds, outflows and jets. They may also give rise to enhanced hydrodynamical 
instabilities, strongly affect the angular momentum evolution of a star throughout its 
lifetime, and they trigger stellar activity with the resulting short-term energetic flares, mass ejections, highly-energetic, accelerated particles, hot coronal plasma in magnetic loops, cool photospheric spots and ionized winds. 
Magnetic fields of cool solar-type stars are ultimately generated via magnetic dynamo processes in the interior of a star, driven by the
interplay of convection and internal differential rotation, thus their strength and topology are a function of spectral type and of stellar
rotation rate (depending on age). 
These dynamo generated fields strongly trigger stellar activity 
and thus generate a significant 
amount of output in UV, FUV, EUV, and X-rays. 
It is this short-wavelength output that, apart from liquid water being present on a planetary surface, 
crucially affects the composition and the physical and chemical 
evolution of upper planetary atmospheres and thus the habitability of a(n) (exo)planet. 

Planets and their host stars form as the final products of an intricate series of physical mechanisms commencing in the galactic interstellar medium. It is during molecular cloud collapse and disk formation, when the initial conditions for later habitable zones are set. The huge diversity in these processes leads to a large range of planetary environments, some of which may be habitable, and in particular the forming and evolving host stars play an extremely important role in controlling the growth of planets and for the evolution of their atmospheres. 
How potentially habitable regions are established is presently rather poorly understood. Apart from the direct starlight allowing for mild climates and liquid water on a planetary surface, many further conditions must be fulfilled and be kept within limits during the violent early years of a star's life, that is rather different from what we encounter in the present-day solar system (e.g. G\"udel et al. 2014).
To address questions on the formation of habitable environments in young stellar/planetary systems, we established the \emph{Pathways to Habitability} project approved by the Austrian Science Foundation (FWF) as a large national key project. The project investigates the astrophysical conditions for planetary habitability addressing the following fundamental questions: \emph{a) What is the evolution history of the gas and dust of protoplanetary disks, and how are water and organic molecules formed and transported to the sites where habitable planets are evolving?} \emph{b) How do magnetic activity, high-energy radiation, the ionized wind, and particles affect the planetary environment and interact with planetary atmospheres?} \emph{c) How do upper planetary atmospheres evolve given the influence of stellar radiation and winds in the presence of a protecting magnetosphere?} and \emph{d) How do dynamics, stellar radiation and winds influence habitable zones in binary systems?}
One of our principal goals is thus to understand the formation and early evolution of habitable environments in young, active stellar surroundings in which high-energy processes and radiation dominate, whereby we will concentrate on the stellar magnetic fields and activity aspect within this proceedings contribution.

\section{Zeeman Doppler Imaging}
A stars' magnetic field is the crucial driver of activity and thus one of the main properties when we attempt to study the influence of a star on its surrounding planet(s). A number of different techniques are currently used to detect and analyze stellar magnetic fields, based on either high-resolution spectroscopy and/or photo- and spectropolarimetry.
The development and the application of the Zeeman Doppler Imaging (ZDI) technique (e.g. Kochukhov et al. 2004, Donati et al. 2006, L\"uftinger et al. 2010a, L\"uftinger et al. 2010b), enables the inversion of a time-series of high-resolution Stokes parameter observations of stellar spectra into surface maps of parameters such as temperature and magnetic field geometry. Built around complex mathematical procedures, this technique has become one of the most powerful astrophysical remote sensing methods. 
Donati (2001), Donati et al. (2006), Piskunov \& Kochukhov (2002), Kochukhov \& Piskunov (2002) and Wade et al. (2001) have described state of the art ZDI codes in detail, which are based on elaborate spectrum synthesis, taking into account all relevant physics of polarized line formation. The code of Kochukhov \& Piskunov (2002) was recently extended towards a new version for temperature and magnetic mapping of the surface structures in cool, active stars (Kochukhov \& Piskunov 2009, Ros\'en \& Kochukhov 2012) including the treatment of molecular opacities.
We would like to mention, however, that ZDI misses possibly present small-scale magnetic flux, whose polarization signatures cancel out, as stated in e.g. Reiners \& Basri 2009. However, the possible neglect of small-scale fields is unlikely to significantly influence the spin-down times and wind configurations calculated from magnetograms derived via ZDI as the large-scale open flux, which is important when studying the influence of stellar fields on surrounding planets, remains unaffected (as found by Lang et al. 2014).   
Due to the development of dedicated novel instrumentation, such as HARPSpol (ESO), ESPaDOnS (CFHT), and NARVAL (TBL, Pic du Midi, France), we now have excellent spectropolarimetric data at our disposal to exploit the full amount of information available from time-resolved observations of Stokes profiles. 


\section{Rotation, Magnetism, and Winds}
Practically all cool, low-mass solar-type stars show magnetic fields comparable to that of our Sun and they can exhibit extraordinary amounts of activity. This activity, in analogy to the solar activity, is likely triggered by dynamo processes in the stellar interior, that generate variable and complex magnetic fields, whose properties strongly correlate with the stellar rotation rate, mass, and age (e.g. Donati \& Landstreet 2009). 
The first successful model for the \emph{solar wind} (Parker 1958) made it clear, that an ionized stellar wind launched from a rotating magnetized star causes a slowing down of the star's rotation rate with time, as angular momentum is transferred from the star by the wind, mainly in the form of stresses in the magnetic field (Weber \& Davis 1967).
Skumanich (1972) was able to prove this observationally, and since then, a large number of studies have shown that although this result is approximately correct under certain conditions, the situation in 'real life' is much more complex. 
Measured rotation periods in young clusters (e.g. Irwin et al. 2008, 2009; Hartman et al. 2010) show that stars start out at the main sequence with a huge two-orders of magnitude spread in rotation rates and as they age and spin-down, at most stellar masses, the rotation rates quickly converge, whereby the spin-down rate is strongly mass dependent.
For stars with solar mass, rotation rates have almost completely converged within the first Gyr (Bouvier et al. 1997, Meibom et al. 2011, Gallet \& Bouvier 2013).

\emph{The Solar Wind}: Our Sun possesses a hot ionized wind that streams outwards in all directions, is in general not isotropic, and can be approximately divided into two distinct components: the slow wind and the fast wind with speeds of approximately 400 km s$^{-1}$ and 800 km s$^{-1}$, respectively. The physical mechanisms that launch and drive the solar wind are not yet well understood (for a review please see Cranmer 2009).
The structure of the solar wind is associated to the structure of the Sun's large scale magnetic field and the speed of the solar wind is closely related to the amount of expansion of the magnetic field between the solar surface and the top of the closed corona (as shown by Wang \& Sheeley, 1990, and Arge \& Pizzo, 2000). 
For a simple axisymmetric dipole magnetic field, i.e. the dominant field structure seen on the Sun at solar minimum, this leads to fast wind coming from the poles and slow wind coming from the equator. 
Complex non-axisymmetric magnetic fields (a configuration seen on the Sun at solar maximum) cause complex distributions of slow and fast wind. 
Thus the solar example clearly shows, that a good understanding of the properties of the stellar wind must be based on a good understanding of the strengths and geometries of stellar magnetic fields.
 
 \emph{Winds of pre-main-sequence stars} are currently only poorly understood. 
Pre-main-sequence stars have coronae that are significantly hotter and several orders of magnitude more luminous in X-rays, than the solar corona.
This suggests that, even though the links between coronal properties and wind properties are not well understood, it can be expected that such stars have winds that are much stronger than the solar wind.
Young stars contract and spin-up as they age and naively one could expect that the accretion of high specific angular momentum material onto the star would lead to accreting stars spinning faster than otherwise expected. It is now firmly established observationally however, that while these stars still possess disks, their rotation rates remain approximately constant. This requires that they lose large amounts of angular momentum (Edwards et al. 1993, Rebull et al. 2004, Gallet \& Bouvier 2013), and whereas the physical mechanisms responsible for this loss of angular momentum are poorly understood, it is clear that they involve star-disk magnetic interactions. The presence of magnetospheric accretion onto the stellar surface is speculated to be a significant source mechanism for stellar winds (Cranmer 2008, Cranmer 2009).

If we want to constrain interactions between exoplanets and the magnetized winds of their host-stars, and to describe the interplanetary medium that surrounds these planets, it is important to adopt more realistic stellar wind models, that also take factors like stellar rotation and complex stellar magnetic field configurations of cool stars into account. 
Thus we have developed a method to determine the characteristics of stellar winds for low-mass main-sequence stars between masses of 0.4 M$_\odot$ and 1.1 M$_\odot$, 
at distances far from the stellar surface, which can then be used as input into models of star-planet interactions (Johnstone et al. 2015a,b, Paper I and II). In Johnstone et al. (2015b), based on 
observationally constrained rotational evolution models, we are able to constrain the mass fluxes in stellar winds and to predict how their properties evolve with time on the main sequence. This way it is possible to couple the stellar wind model developed in Paper I, with the rotational evolution model that we developed in Paper II, and to predict, how stellar wind properties evolve on the main-sequence for a given range of stellar masses. 
In Figure\,1 we show, as an example, the slow and fast components of the solar wind velocity against radial distance for the
solar wind models derived by us (Johnstone et al. 2015b). 
We estimate that for the young Sun, the solar wind likely had a mass loss rate that was an order of magnitude higher than that of the current solar wind. 

As a result  of the spread in rotation rates, young stars exhibit a large range of wind properties at a given age. This spread disappears as the stars age. We therefore conclude that there is a significant uncertainty in our knowledge of the evolution of the solar wind, due to the lack of theoretical or observational knowledge of stellar winds, and also due to the above mentioned large spread in rotation rates at young ages. Taking the sensitivity of planetary atmospheres to stellar wind and radiation conditions into account, these uncertainties are essential for our understanding of the evolution of habitable planetary environments. 
To learn from the history of our Earth's (now) habitable atmosphere, it is crucial to constrain, about where the Sun might have been in the distribution of rotation rates at 100 Myr.
Although we detect a specific dominance of the slowly rotating evolutionary track at this age, there is a significant possibility that the Sun was a rapid rotator at 100 Myr. This indeciciveness is important as these scenarios correspond to radically different histories for the 
Sun's magnetic activity and as a consequence to radically different levels of high energy
radiation and strengths of its wind. From the stellar point of view, these uncertainties are crucial for studies investigating the evolution of planetary atmospheres, as their upper atmospheres are extremely sensitive to the level of high-energy radiation from the central star (Tian et al. 2008; Lammer et al. 2010).
For more detailed discussions we here refer to our Papers I and II: Johnstone et al. 2015\,a\&b. 

\begin{figure}[t]
\begin{center}
\includegraphics[scale=0.35]{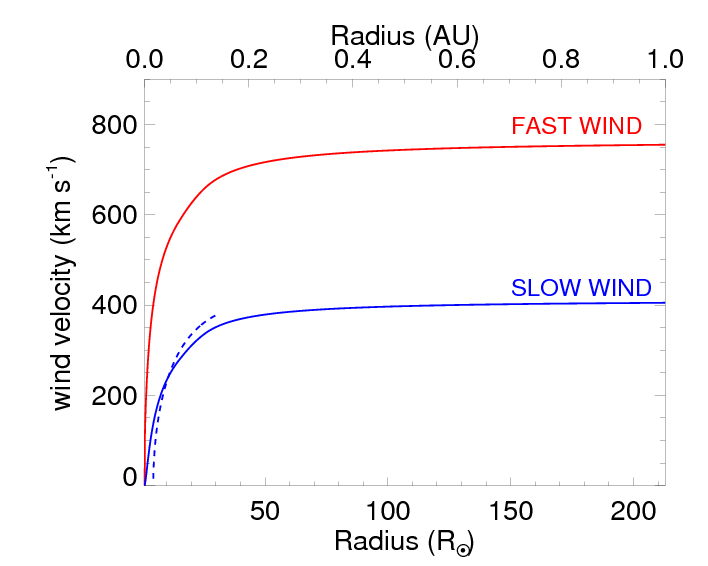}
\caption{Wind velocity against radial distance from the Sun for the
solar wind models derived by Johnstone et al. (2015). The blue (bottom) and red (top) 
lines show the models for the slow and fast components of the solar wind
respectively. The dashed (blue) line shows the observational constraints on
the slow wind acceleration derived by Sheeley et al. (1997).
}
\end{center}
\end{figure} 

The magnetism of cool, low-mass stars can be significantly different from the solar one in topology as well as in intensity. ZDI reconstructions of stellar large-scale magnetic fields allow us to obtain even more realistic models of magnetized stellar winds, incorporating magnetic maps, determined from spectropolarimetric observations, into numerical simulations (Vidotto et al. 2012, Vidotto et al. 2014, Jardine et al. 2013, Llama et al. 2013). The radial component of the magnetic field $B_r$ is used as a boundary condition in these models, and held fixed throughout the simulation run. While the simulations evolve in time, both the wind and the magnetic field lines interact with each other and the solution (obtained self-consistently) is found when the system reaches steady state in the reference frame corotating with the star.  

\emph{Case studies}: \emph{$\tau$~Boo} is one of the currently most outstanding planet hosting stars. $\tau$ Boo hosts a giant planet orbiting very close in and also exhibits a large scale magnetic field that shows cyclical polarity flips (Catala et al. 2007, Donati et al. 2008a, Fares et al. 2009, Fares et al. 2013). Numerical simulations of the stellar wind of this star were performed by Vidotto et al. (2012), based on the surface magnetic maps (see above) as boundary conditions for the stellar wind simulations. They found
that the outflowing wind is directly influenced by variations of the stellar magnetic field during the cycle. 
\\
\emph{Magnetic moment and plasma environment of 209458b}: 
Indirect hints for the properties of a stellar wind and the planetary magnetic moment are explored by Kislyakova et al. (2014) as part of our research network (PatH, see Introduction): they investigate the causes of strong absorption in both the blue and red wings of the stellar Lyman-$\alpha$ line revealed by line transit observations of HD 209458b, obtained by the HST. The absorption is interpreted as hydrogen atoms escaping at high velocities from the exosphere of the planet. Sources proposed for this effect are natural spectral line broadening, the acceleration by stellar radiation pressure, 
or the charge exchange with the stellar wind. Kislyakova et al. (2014) reproduce the observations by means of
modeling that includes all aforementioned processes and their results support a stellar wind with a velocity of about 400 km/s during the time of observation and a planetary
magnetic moment of $\approx{1.6 \times 10^{26}}$ amp\`eres per square meter.
\section{Summarizing Remarks}
It is well accepted that stellar magnetic fields play an important role in the evolution of planetary system, though still poorly constrained from the observational point of view. 
Present studies show that magnetic field topologies are often significantly more complex than a simple dipole or quadrupole. Additionally, the long-term evolution of stellar magnetic fields needs to be taken into account, as, during their evolution history, fields considerably change their properties, plus, 
not only the Sun, but also other stars have been observed to show the presence of magnetic field cycles. It is therefore crucial, when studying the influence of host stars on their surrounding planets, to continue obtaining time-resolved spectropolarimetric observations, applying the ZDI technique on spectropolarimetric data, and to benefit from the resulting maps of field geometries as input for studies on wind modelling and star-planet interactions.  

\begin{acknowledgement}
TL acknowledges funding of the Austrian FFG within ASAP11; TL, MG and CPJ acknowledge support by the FWF NFN projects S11601-N16 and S116 604-N16.
\end{acknowledgement}

\end{document}